\documentclass[preprint,10pt,prl,onecolumn]{revtex4}
\usepackage{amssymb}

\input{tcilatex}

\begin{document}

\title{Photon-exchange effects on photon-pair transmission}
\author{K.J. Resch, G.G. Lapaire, J.S. Lundeen, J.E. Sipe, and A.M. Steinberg%
}
\affiliation{Department of Physics, University of Toronto\\
60 St. George Street, Toronto ON M5S 1A7\\
CANADA}

\begin{abstract}
It has been proposed that photon-exchange effects in atom-photon
interactions could lead to greatly enhanced optical nonlinearities. These
might have widespread application (e.g. quantum information). \ Here we
demonstrate experimentally that such exchange effects can indeed enhance the
probability of real absorption of photon pairs. \ Using nonclassical pairs
of photons with variable time separation, we observe a maximum suppression
of pair transmission by at least 5\% with respect to the result for
uncorrelated photons.
\end{abstract}

\maketitle

Exchange effects play a rich and central role in quantum physics, and are
linked with phenomena as diverse as the Pauli exclusion principle and Bose
condensation. For both fermions and bosons they result in a modification of
the effect of inter-particle interaction from what would be predicted for
distinguishable particles, and can often be thought of as leading to a new
effective interaction between indistinguishable particles. Photons are
bosons, and are usually thought to interact with each other, \textit{via}
their coupling with material media, very weakly. For example, the usual Kerr
nonlinearity, which is well-known to grow linearly with the atom density in
the interaction region \cite{chi3} rarely produces phase shifts larger than $%
10^{-10}$ for a pair of photons. \ This can be enhanced using schemes
employing cavity QED \cite{qed}, electromagnetically-induced transparency
and slow light \cite{eit}, or interference-based nonlinearities \cite%
{switchnlo}. \ However, the technical difficulties that plague these
protocols have led many who seek an effective photon-photon interaction for
applications in quantum computing and information processing to turn instead
to the use of linear optics and conditional detection \cite%
{klm,otherklmlikeschemes} to simulate such an effect. Could an exchange
effect instead be leveraged to lead to an effective photon nonlinearity at
the few photon level?

Six years ago Jim Franson \cite{franson,otherfranson} argued that indeed
photon-exchange interactions in an atomic system might give rise to a very
strong effective nonlinearity. The process requires a pair of photons at
frequencies $\omega _{1}$ and $\omega _{2}$, and a pair of atoms $A$ and $B$%
. Exchange would occur when atom $A$ nonresonantly `absorbs' photon $1$ and
emits photon $2$, while atom $B$ absorbs photon $2$ and emits photon $1$.
Franson predicted that this exchange-based nonlinear effect would grow as
the square of the atom density and could be significant at the two-photon
level in a sufficiently dense sample. This proposed effect has been the
subject of significant controversy \cite{antifranson}, and has not yet been
verified in the laboratory.

In this letter we present both a theoretical motivation and experimental
realization of a simpler photon-exchange effect, one that involves \textit{%
real }transitions in matter rather than \textit{virtual }transitions.
Consider the two-photon absorption probability for like-polarized photons
travelling in the same direction in a one-dimensional system. This can
easily be calculated in perturbation theory. We consider a two-photon state
initially described as: 
\begin{equation}
\left| A,B\right\rangle =\mathcal{N}(A,B)\iint d\omega d\omega ^{\prime
}f_{A}(\omega )f_{B}(\omega ^{\prime })a^{\dagger }(\omega )a^{\dagger
}(\omega ^{\prime })\left| 0\right\rangle ,  \label{twophotonstate}
\end{equation}%
where $a^{\dagger }(\omega )$ is a raising operator for a photon of
frequency $\omega $, $f_{A(B)}(\omega )$ is the frequency amplitude function
for an individual photon $A(B)$, and $\mathcal{N}(A,B)=\left( 1+\left| \int
f_{A}^{\ast }(\omega )f_{B}(\omega )d\omega \right| ^{2}\right) ^{-1/2}$ is
a normalization constant, $f_{A(B)}(\omega )$ are two amplitude functions
for frequency $\omega $, and $a^{\dagger }(\omega )$ is the raising operator
for a photon of frequency $\omega $. \ Under these conditions, the
probability of absorbing both photons, $P_{AB},$ can be expressed in terms
of the single-photon absorption probabilities $P_{A}$ and $P_{B}$: 
\begin{equation}
P_{AB}=P_{A}P_{B}\left( \frac{1+\xi _{AB}}{1+\upsilon _{AB}}\right) ,
\end{equation}%
where $\upsilon _{AB}$ is the square of the overlap integral, $\left| \int
d\omega f_{A}^{\ast }(\omega )f_{B}(\omega )\right| ^{2}\leqslant 1$, and 
\begin{equation}
\xi _{AB}=\frac{\left| \int d\omega g(\omega )f_{A}^{\ast }(\omega
)f_{B}(\omega )\right| ^{2}}{\left[ \int d\omega g(\omega )\left|
f_{A}(\omega )\right| ^{2}\right] \left[ \int d\omega g(\omega )\left|
f_{B}(\omega )\right| ^{2}\right] },
\end{equation}%
where $g(\omega )$ is the absorption spectrum of the medium. \ For
independent absorption events we expect $P_{AB}=P_{A}P_{B};$ therefore
correlated absorption probabilities come from cases where $\upsilon
_{AB}\neq $ $\xi _{AB}.$ \ Such a case can be set up if the two photons are
separated in time, but pass through a medium with a narrow absorption
feature of width, $\Delta \omega _{a}$. \ Since the photons do not overlap, $%
\upsilon _{AB}\approx 0$. Provided that the absorber has a coherence time, $%
\tau \equiv 1/\Delta \omega _{a}$, longer than the delay between the
photons, we may have $\xi _{AB}\neq 0.$ \ In other words, if the photons are
distinguishable before absorption, but become (mostly) indistinguishable if
absorbed, we may have $P_{AB}>P_{A}P_{B}$ -- an enhancement of the
two-photon absorption. \ In cases where the photon delays are much longer
than the coherence time of the absorber, or when the photons are perfectly
overlapped, $P_{AB}=P_{A}P_{B}$, as expected in the absence of any nonlinear
effects$.$

It is possible to move beyond a perturbative approach with a
photon-wavefunction picture, even for two-photon states $\left|
A,B\right\rangle $ which are not restricted to one-dimension (as in Eq. \ref%
{twophotonstate}). \ But in that one-dimensional limit (as in Eq. \ref%
{twophotonstate}), in the absence of any absorber the coincidence detection
rate of two ideal detectors at positions $z_{1}$, $z_{2}$ and at times $%
t_{1} $, $t_{2}$ is proportional to,%
\begin{equation}
w^{(2)}(z_{1},z_{2},t_{1},t_{2})=\left| \mathcal{N}(A,B)\right| ^{2}\left[ 
\begin{array}{c}
|\psi _{A}(z_{1},t_{1})|^{2}|\psi _{B}(z_{2},t_{2})|^{2}+|\psi
_{A}(z_{2},t_{2})|^{2}|\psi _{B}(z_{1},t_{1})|^{2}{} \\ 
+\psi _{A}^{\ast }(z_{1},t_{1})\psi _{B}(z_{1},t_{1})\psi _{B}^{\ast
}(z_{2},t_{2})\psi _{A}(z_{2},t_{2}){} \\ 
+\psi _{A}^{\ast }(z_{2},t_{2})\psi _{B}(z_{2},t_{2})\psi _{B}^{\ast
}(z_{1},t_{1})\psi _{A}(z_{1},t_{1}){}.%
\end{array}%
\right] .  \label{secondorder}
\end{equation}%
The first-order photon wavefunctions, $\psi _{A(B)}(z,t)$, satisfy $%
E^{+}(z,t)\left| A(B)\right\rangle =\psi _{A(B)}(z,t)\left| 0\right\rangle $%
, where $E^{+}(z,t)$ is the positive frequency component of the electric
field operator, and $\left| A(B)\right\rangle =\int d\omega f_{A(B)}(\omega
)a^{\dagger }(\omega )\left| 0\right\rangle $ is a single-photon state. \
The last two terms on the right-hand side of Eq. \ref{secondorder} are the
exchange terms. In the absence of these terms, the coincidence detection
rate is proportional to the sum of the product of the individual photon
detection rates at either detector, which is characteristic of independent
detection events. Applying this wavefunction description to a Hong-Ou-Mandel
interferometer \cite{hom}, one can show that the Hong-Ou-Mandel dip is
simply a manifestation of photon exchange. \ Considering now the presence of
an absorbing medium, the photon wavefunctions allow one to see the
underlying physics in which, in this case, destructive interference occurs
between the detection events. \ The wavefunctions, perhaps initially
nonoverlapping before passage throught the absorbing medium, may no longer
be after their passage through it. \ Thus there can be interference effects
in the subsequent detection, and a corresponding reduction in the two-photon
transmission probability below the uncorrelated-absorption prediction. \
This is the experimental signature we seek.

To investigate the exchange-induced suppression of two-photon transmission
we used the basic system sketched in Fig. 1. \ There are essentially three
parts to both the experiment and the theory: state preparation, evolution,
and detection. \ The quantum state of interest is a normalized two-photon
state where both photons have the same polarization, same spectrum and a
variable inter-photon time delay. \ This state evolves as it passes through
an absorptive medium which has some resonant absorption feature in the
centre of the photon spectrum. \ Finally, the photon pairs that emerge from
the medium are counted.

\label{experiment}

The two-photon states were created in the setup shown in Fig. 2. \ The setup
is similar to a polarization-based Hong-Ou-Mandel\ (HOM) interferometer \cite%
{hom,sasha} whereby the desired two-photon state is post-selected through
the detection of a photon pair. \ Specifically, we used a collinear type-II
phase-matched parametric down-conversion source (0.1-mm thick BBO) pumped by
the second harmonic of a Ti:Sapphire laser. \ The second harmonic was
centred at 405nm (with a bandwidth of 7nm FWHM) and created pairs of photons
each with centre wavelengths of 810nm. \ The photon pairs exit from the
source such that one has vertical polarization and the other horizontal. \
We control the relative time delay between the photons by passing them
through a modified Babinet compensator. \ After the variable delay, a
polarizer is placed in the photons' path at 45$^{\circ }.$ With the
polarizer in the system, any photon pairs transmitted through this polarizer
are polarized at 45$^{\circ }$, with a time separation determined by the
Babinet. \ This serves as the source of two-photon states required by the
theoretical work. \ From HOM\ interference \cite{hom}, we expect an increase
in the number of photon pairs created near zero delay as the photons tend to
pair up. \ We measure the rate of photon-pair production for each delay, in
the absence of any absorber, and use this to normalize our subsequent
absorption experiment.\ 

The theory also requires an absorber with an absorption feature narrower
than the bandwidth of the photon pairs. \ From our down-conversion source,
we obtain photon pairs with a FWHM power spectrum of over $100$nm. \ We used
a dielectric interference filter (CVI F10-810-4-1.00) as an effective
absorbing medium. \ The back of the filter was blacked-out, and so any light
transmitted through it was discarded. \ The reflectivity of the filter shows
a 10 nm wide dip centred at 810nm. \ The reflected light rhus plays the role
of light nominally transmitted through an absorber with a 10nm absorption
feature, and we refer to it as such below. \ It should be noted that a
gaseous atomic sample could be used if one used a narrower bandwidth
down-conversion source \cite{oushapiro}.

We began by examining the reflection from a broadband dielectric mirror in
order to measure the number of reflected photon pairs as a function of delay
using a cascaded pair of SPCMs (Perkin Elmer SPCM-AQR-13) \cite{4-photon}. \
This data set shows any changes in the efficiency of two-photon state
production, which will be divided out.\ \ Then the interference filter (with
the blacked-out back) is placed directly in front of a broadband dielectric
mirror. \ With the filter in place, we measured the number of remaining
photon pairs in the reflection. \ We monitored the detectors' singles rates
and their coincidence rate in the experiment.

Fig. 3 shows the rates of photon-pair detection (solid circles) and singles
rate (small open diamonds) at one of the detectors as a function of the time
delay between the photons. \ Fig. 3a shows the data taken with the broadband
mirror in place. \ It clearly shows that the two-photon state preparation
becomes much more efficient at zero delay. \ This increased pairing is due
to HOM interference \cite{hom}. \ In our experimental results, the rate of
photon pair production at zero delay is $55\%$ larger than that at large
time delays, whereas perfect HOM interference leads to a doubling of the
rate. \ As one would expect from a HOM\ interferometer with low collection
or detection efficiencies, the singles rate is featureless at the $1\%$
level as a function of the time delay between the photons \cite{boga}. \
Fitting the data in Fig. 3a under the assumption of identical gaussian power
spectra for the two photons yielded a FWHM\ of 129nm. \ 

Fig. 3b shows the data taken with the interference filter in place. \ The
most striking difference from the previous figure is the drop in the number
of photon pairs detected at a delay of approximately $\pm 10$ fs. \ There is
a second, more subtle, difference in that the number of photon pairs at zero
delay is enhanced by only $42\%$ over the rate at large time delays with the
filter in place. \ The singles rate in Fig. 3b also shows no dependence on
the time delay at the $1\%$ level.

The ratio of the data in Fig. 3b to Fig. 3a is shown in Fig. 4 as solid
circles. \ This ratio normalizes the data for changes in the production
efficiency of two-photon states and shows the photon pair reflection
probability given a two-photon input. \ To reduce the noise on the data
points a 5-point average was taken. \ We compare the observed rate of photon
pair transmission at $\pm 10$ fs to long- and zero-time delays. \ The
normalized photon-pair detection ratio at $-10$ fs is $15\%$ less than that
at long delays and $5\%$ less than that at zero; the ratio at $+10$ fs is
about $17\%$ less than at long delays and $7\%$ less than that at zero. \
The asymmetry in the data may be due to dispersion effects in the nonlinear
crystal \cite{dipshape}. \ Fig. 4 also shows the theoretical predictions for
the coincidence rate ratio as a function of the time delay between the
photons in two different regimes using gaussian spectra and a gaussian
absorption feature. \ The two different curves correspond to photon spectra
that are initially uncorrelated (solid curve), as in Eq. 1, and where the
sum of the photon energies is a constant (dotted curve). \ The first regime
describes the light produced from a down-conversion source with a broadband
pump laser after sufficiently narrow bandpass filtering, whereas the second
describes down-conversion created by a CW laser \cite{mete,grice}. \ The
curves are both scaled to their coincidence rate at very large time delays.
\ \ The drop in the ratio measured at zero delay which is not predicted by
the theory. \ In our theoretical description, the photons are perfectly
indistinguishable at zero time delay. \ Thus, passing through the absorbing
medium can do nothing to change their degree of distinguishability and the
rate of photon-pair absorption is that of uncorrelated events. \ In our
experiment, however, the photons are still partially distinguishable even at
zero delay due to imperfect spectral and spatial mode-matching. \ When some
light is absorbed by the medium, the overlap between the photon
wavefunctions are changed and this leads to an additional effect at zero
delay. \ Photon-exchange terms are responsible for at least the $5\%$ change
in the photon pair transmission through our absorbing media at $\pm 10$ fs.
\ 

In both situations, the enhanced region of photon absorption occurs at
approximately $\pm 10$ fs, in good agreement with experiment. \ The $5\%$
suppression in the rate of photon-pair transmission is on the order of that
predicted by our theory, but a more accurate comparison will require better
data or a more complete theory that includes imperfect spatial- and
spectral-mode matching. \ The shape of our experimental curve is in better
agreement with the theoretical curve with no frequency correlations where
the transmission reaches its long delay value at about $\pm 20$ fs. \ The
theory for perfect frequency correlations shows suppressed transmission over
much longer times. \ In our experiment we are actually between these two
extreme regimes; our ultrafast-pump laser has a bandwidth of about $7$ nm. \
From preliminary calculations, we have found that by accounting for partial
frequency correlations, suppression in photon pair transmission can be
enhanced over the case with no correlations without significantly changing
the shape of the curve. \ It is clear from the theory that these
correlations can greatly influence the photon-pair transmission probability.
\ Such a striking dependence on these correlations makes this technique
useful for measuring them. \ 

In summary, we have observed that photon-exchange effects can give rise to
nonlinear behaviour on the pair-transmission probability. \ Like-polarized
photon pairs with a variable inter-particle delay have been shown to exhibit
suppressed two-photon transmission through a linearly absorbing medium -- an
experimental signature of exchange enhancement of photon-pair absorption. \
This suppression occurs for delays that are longer than the photons'
coherence times but shorter than the coherence time of the absorber. \
Further work is needed to understand the limits on the applicability of such
exchange effects to nonlinear optics and quantum information.

This work was funded by Photonics Research Ontario, NSERC, and the U.S. Air
Force Office of Scientific Research (F49620-01-1-0468).\

\textbf{Fig. 1. \ The system of interest. \ A two-photon state with a
variable time delay between the like-polarized photons impinges on an
absorbing medium. \ The light that passes through the absorption region is
detected by a photon-pair detector. \ The probability of absorbing a photon
pair can depend very strongly on the delay.}

\textbf{Fig. 2. \ The experimental setup. \ The state preparation is
accomplished using the output of a polarization-based HOM interferometer %
\cite{hom,sasha}. \ BBO is a }$\beta $\textbf{-barium borate nonlinear
crystal phase-matched for type-II down-conversion; PBS is a polarizing
beamsplitter; SPCMs are single photon counters; Pol. is a polarizer. \ The
pump laser is separated from the down-conversion beams using a fused silica
prism (not shown). \ The two-photon state is prepared conditioned on
successful post-selection of a photon pair after the polarizer. \ Once
prepared in the right quantum state, the light reflects from either a
broadband dielectric mirror or a removable interference filter (I.F.) (CVI
F10-810-1.00-4) (to simulate a medium with a very broad absorption line). \
The two-photon state is prepared conditionally, when both photons are
transmitted through the 45-degree polarizer. \ This 45-degree polarized
light is then split at a polarizing beamsplitter which has a single-photon
detector in each output. \ The PBS\ acts as a 50/50 beamsplitter on the 45}$%
^{\circ }$\textbf{\ polarized photons which allows the pair of SPCMs to act
as a two-photon detector.}

\textbf{Fig. 3. Experimental data. \ The coincidence rate (solid circles)
and a singles rate (small open diamonds) as a function of the inter-photon
delay are shown in the case where a) the `absorber' (interference filter) is
removed and when b) the `absorber' is in place.}

\textbf{Fig. 4. \ Experimental and theoretical normalized ratio of
coincidence rates. \ Plotted is the normalized ratio of the coincidence
rates of Fig. 3b to Fig. 3a. \ \ The maximum drop in the coincidence rate
ratio occurs for delays of }$\pm 10$\textbf{\ fs and is at least 5\% less
than the rate at zero delay. \ The data have been normalized to the average
rate over the delay, }$\left| \tau \right| \geq 20$\textbf{fs. \ The
theoretical predictions are also shown. \ \ The case where there are
initially no frequency correlations between the two photons is shown as a
solid curve. \ The case where the frequencies sum to a well-defined value is
shown as a dotted curve.}

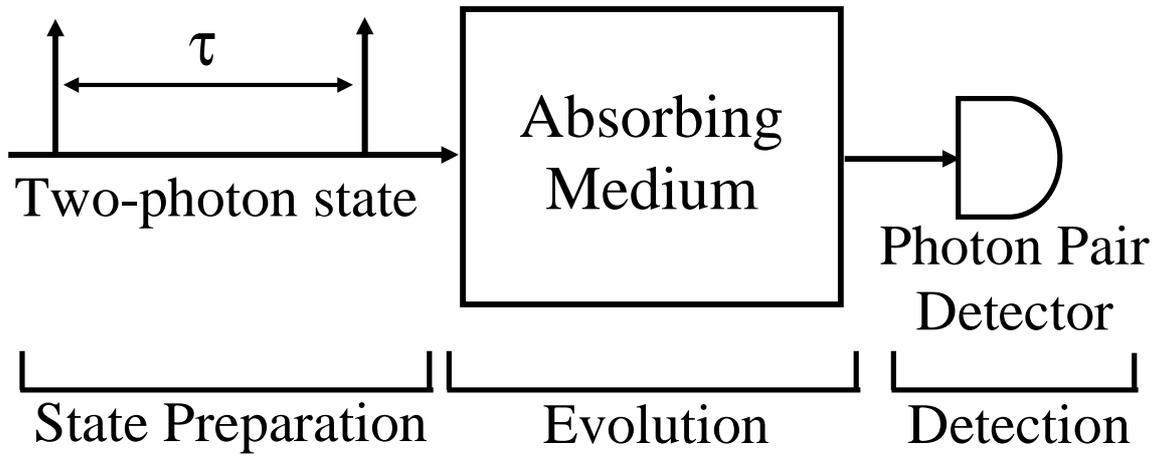
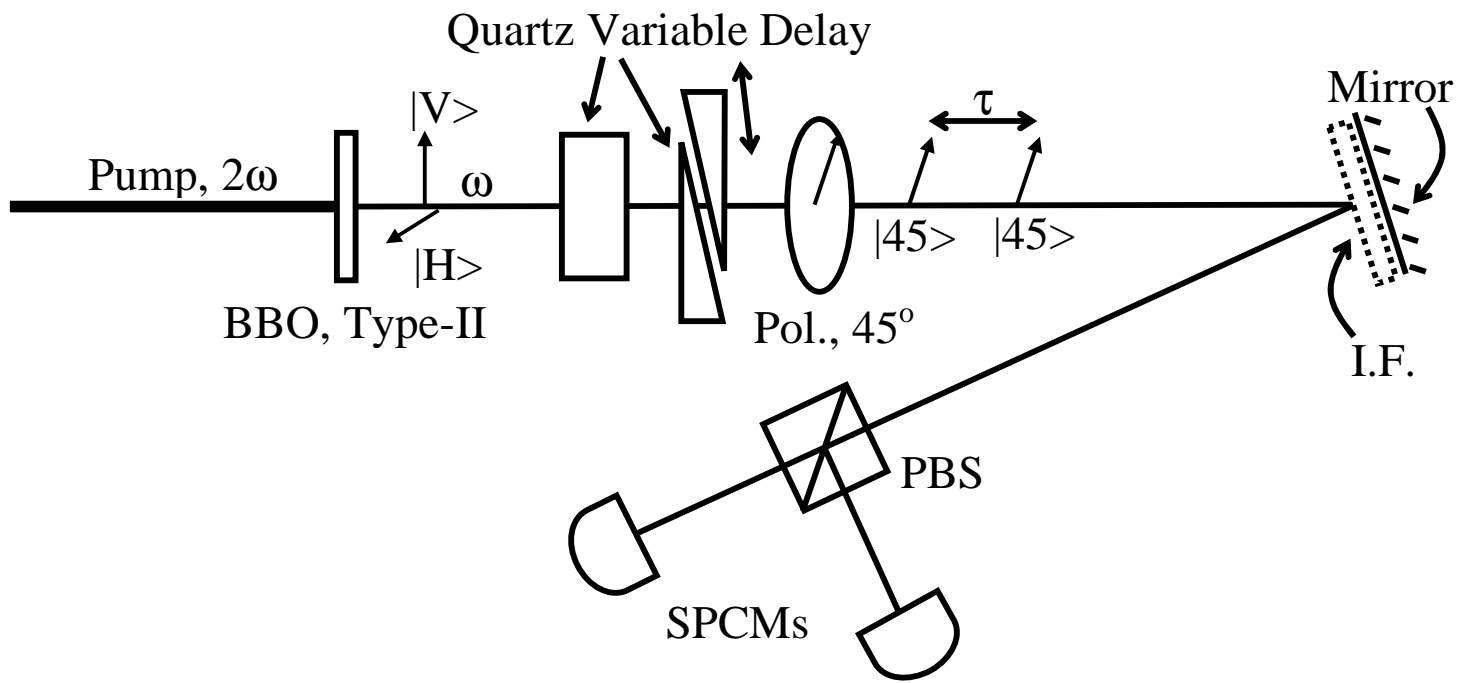
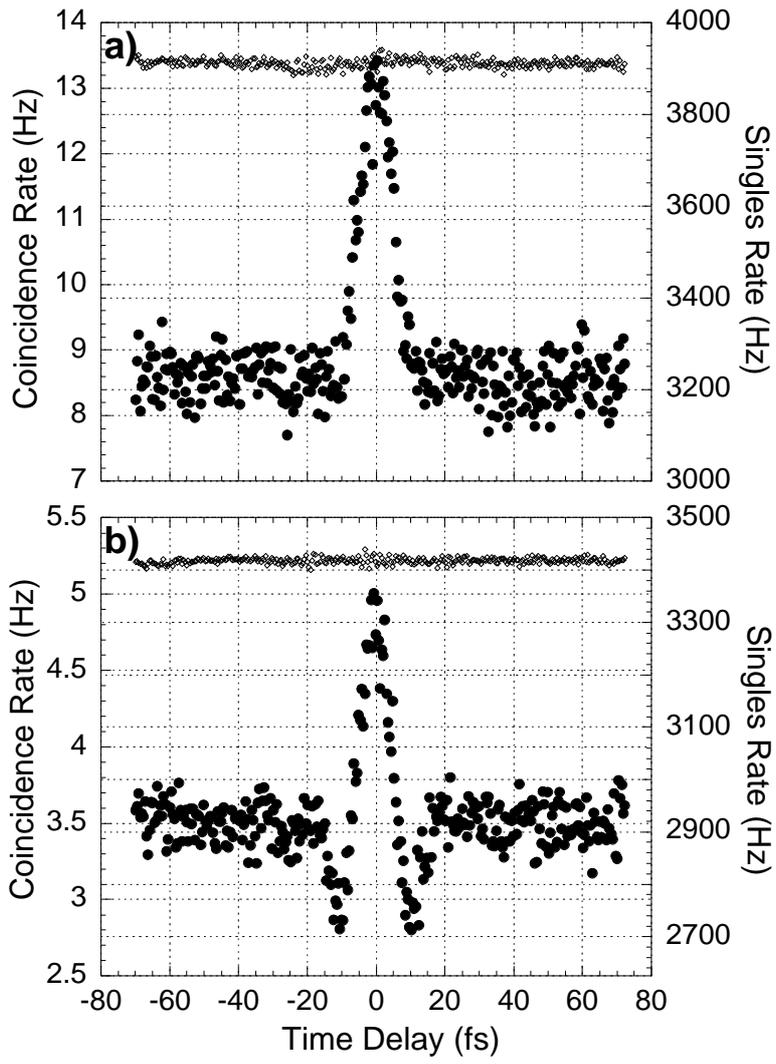
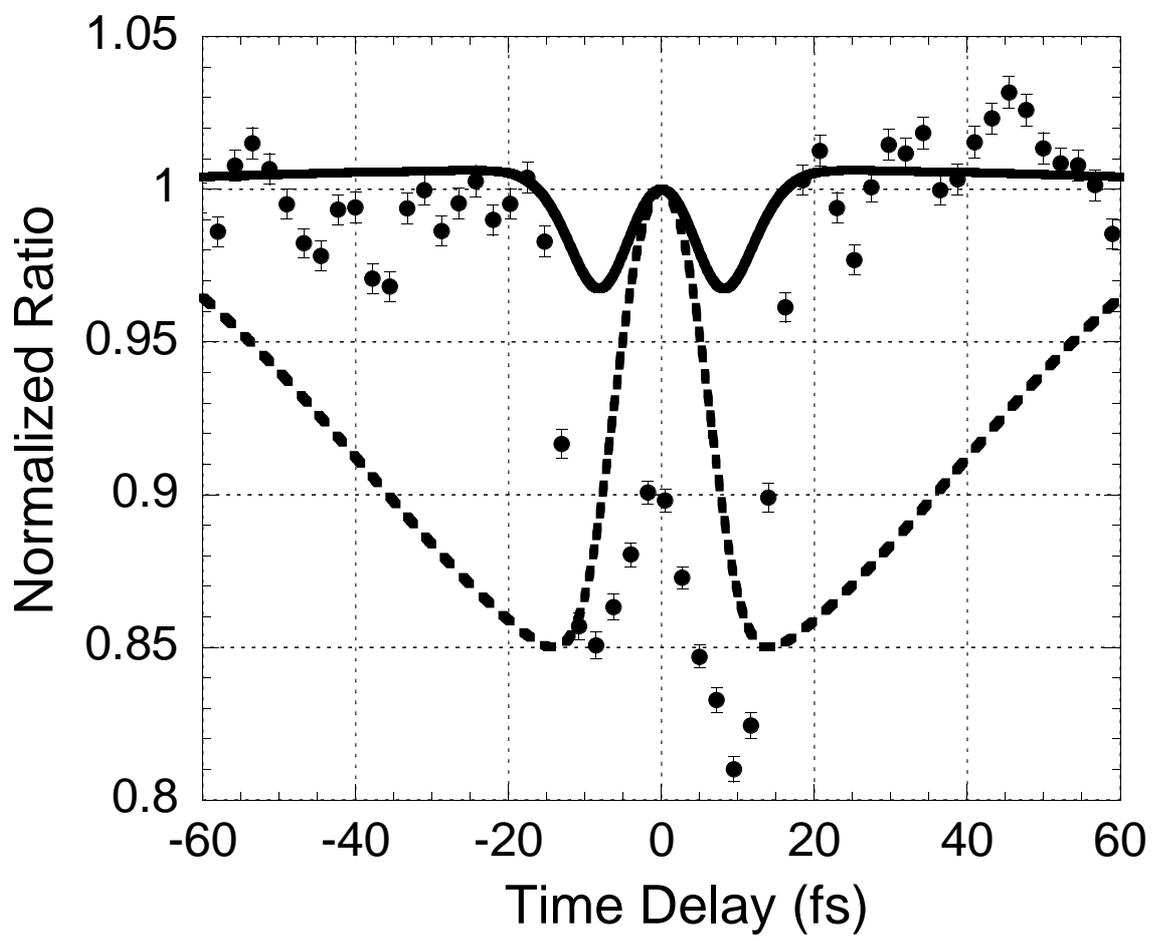
\end{document}